\documentclass[preprint]{elsarticle}

\usepackage{amssymb}
\usepackage{amsmath}
\usepackage{siunitx}
\usepackage{enumitem}
\usepackage[hyphens]{url}
\usepackage{hyperref}
\usepackage{booktabs} 
\usepackage{makecell}
\usepackage{subcaption}
\usepackage{mathtools}
\usepackage{breakurl}

\begin{document}

\begin{frontmatter}




\title{Mapping intratumoral heterogeneity through PET-derived washout and deep learning after proton therapy}




\author[ucm,clinico]{Pablo Cabrales\corref{corr}} 
\ead{pcabrale@ucm.es}
\author[harvard,uc3m]{David~Izquierdo-García}
\author[harvard]{Víctor~V.~Onecha}
\author[ucm,clinico]{Mailyn~Pérez-Liva}
\author[ucm,clinico]{Luis~Mario~Fraile}
\author[ucm,clinico]{José~Manuel~Udías}
\author[ucm,clinico]{Joaquin~L.~Herraiz}

\affiliation[ucm]{
  organization = {Grupo de Física Nuclear, EMFTEL \& IPARCOS, Universidad Complutense de Madrid, CEI Moncloa},
  postcode     = {28040},
  state        = {Madrid},
  country      = {Spain}
}

\affiliation[clinico]{
  organization = {Health Research Institute (Instituto de Investigación Sanitaria) of the Hospital Clínico San Carlos},
  postcode     = {28040},
  state        = {Madrid},
  country      = {Spain}
}

\affiliation[harvard]{
  organization = {Massachusetts General Hospital and Harvard Medical School},
  postcode     = {02114},
  city         = {Boston},
  state        = {MA},
  country      = {USA}
}

\affiliation[uc3m]{
  organization = {Bioengineering Department, Universidad Carlos III de Madrid},
  postcode     = {28911},
  state        = {Madrid},
  country      = {Spain}
}

\cortext[corr]{Corresponding author:}

\begin{abstract}

The distribution of produced isotopes during proton therapy can be imaged with Positron Emission Tomography (PET) to verify dose delivery. However, biological washout, driven by tissue-dependent processes such as perfusion and cellular metabolism, reduces PET signal-to-noise ratio (SNR) and limits quantitative analysis. In this work, we propose an uncertainty-aware deep learning framework to improve the estimation of washout parameters in post-proton therapy PET, not only enabling accurate correction for washout effects, but also mapping intratumoral heterogeneity as a surrogate marker of tumor status and treatment response. We trained the models on Monte Carlo–simulated data from eight head-and-neck cancer patients, and tested them on four additional head-and-neck and one liver patient. Each patient was represented by 75 digital twins with distinct tumoral washout dynamics and imaged 15 minutes after treatment, when slow washout components dominate. We also introduced \textit{washed-out} maps, quantifying the contribution of medium and fast washout components to the loss in activity between the end of treatment and the start of PET imaging. Trained models significantly improved resolution and accuracy, reducing average absolute errors by 60\% and 28\% for washout rate and washed-out maps, respectively. For intratumoral regions as small as 5 mL, errors predominantly fell below thresholds for differentiating vascular status, and the models generalized across anatomical areas and acquisition delays. This study shows the potential of deep learning in post-proton therapy PET to non-invasively map washout kinetics and reveal intratumoral heterogeneity, supporting dose verification, tumor characterization, and treatment personalization. The framework is available at \href{https://github.com/pcabrales/ppw.git}{https://github.com/pcabrales/ppw.git}.


\end{abstract}



\begin{keyword}
Biological Washout \sep Positron Emission Tomography (PET) \sep Deep Learning \sep Proton Therapy \sep Intratumoral Heterogeneity \sep Monte Carlo Simulation \sep Digital Twins



\end{keyword}

\end{frontmatter}

\section{Introduction}


In proton therapy~\cite{lower-integral-dose}, the accuracy of dose delivery is limited by factors such as patient positioning errors, organ motion, and uncertainties in treatment planning parameters, resulting in unintended damage to surrounding healthy tissue and organs at risk~\cite{uncertainties-list, range-uncertainties-paganetti}. Therefore, verifying the delivered dose is essential to detect deviations, enable treatment adaptations, and improve precision~\cite{safety-margins, parodi2015Vision}.


Since protons deposit most of their energy near the end of their range and do not exit the patient~\cite{PT-physics}, dose verification primarily relies on detecting secondary emissions~\cite{verification-review}. Among these methods, the most extensively studied involves the use of PET imaging, which detects positron-emitting isotopes generated by nuclear interactions as protons traverse the patient's tissues. The main produced isotopes, or positron emitters, include $^{11}$C, $^{13}$N, $^{15}$O, and $^{38}$K~\cite{pet-verification-paganetti-fakhri}. Approaches to activate extrinsic contrast agents and produce alternative positron emitters have also been proposed~\cite{isotopes-contrast}.


PET-based verification 
can be performed in two different ways: (1)~using purpose-built PET scanners during irradiation (\textit{in-beam} approach), or (2)~moving the patient to a commercial scanner shortly after treatment (\textit{offline} approach)~\cite{on-vs-offline-parodi}. The offline approach can be readily implemented in many proton therapy centers, as it requires only access to a conventional PET scanner rather than specialized, high-cost systems. Nevertheless, its effectiveness is constrained by low PET count statistics due to intrinsically low activation yield~\cite{review-pet-pt} and biological washout of positron emitters~\cite{offline-bauer}.

Biological washout is a complex, dynamic, and tissue-specific process in which positron emitters are displaced from their original production site~\cite{washout-tomitani}. This effect further degrades the signal-to-noise ratio (SNR) in post-proton therapy PET images, and it does so in a tissue-dependent, spatially heterogeneous manner. In living tissues, positron emitters can be displaced as they are formed at, or bound to, molecular compounds and intervene in biological processes like cellular metabolism and perfusion. Consequently, tissue-specific microvascular architecture and metabolic activity heavily influence biological washout, making it difficult to model comprehensively~\cite{parodi2015Vision}.

\begin{figure}[h]
    \centering
    \includegraphics[width=\linewidth]{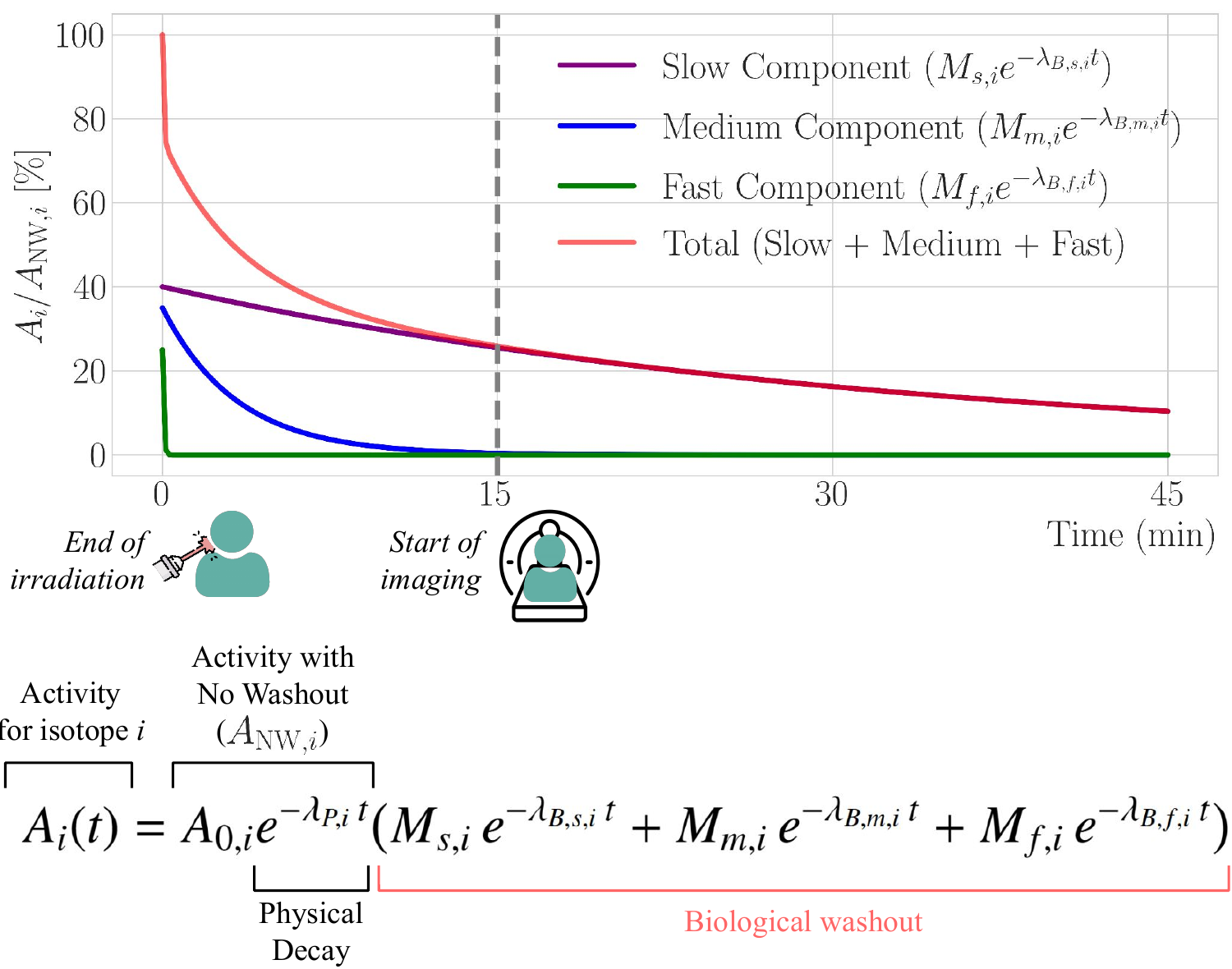}
    \caption{Schematic representation of the reduction in activity due to biological washout for a positron-emitting isotope ($i$). The model includes three components with typical biological washout rates for $^{11}\text{C}$: $\lambda_{B,s,i} = 0.03~\text{min}^{-1}$ (slow), $\lambda_{B,m,i} = 0.30~\text{min}^{-1}$ (medium), and $\lambda_{B,f,i} = 20~\text{min}^{-1}$ (fast). The corresponding component fractions at the end of the irradiation, also representative of typical values, are $M_{s,i} = 0.40$, $M_{m,i} = 0.30$, and $M_{f,i} = 0.25$, respectively. The washout kinetics of the other isotopes can also be characterized by a three-component model. In an offline PET imaging approach where acquisition begins 15 minutes post-treatment, the measured activity signal is dominated by the slow component of $^{11}\text{C}$ washout. In this study, washout rate maps provide voxel-wise quantification of the slow biological washout rate of $^{11}\text{C}$ ($\lambda_{B,s,i}$). In contrast, \textit{washed-out} maps provide voxel-wise quantification of $A_i/A_{\text{NW}, i}$ at the start of imaging, which represents the fraction of activity remaining after biological washout and primarily reflects the contributions of the medium and fast washout components.
    }
    \label{fig:washout-decay}
\end{figure}

In proton therapy, biological washout is commonly approximated by a one-compartment kinetic model, where the reduction in PET signal within the treated tissue is described as a decay process involving three components with different biological washout, or clearance, rates: fast, medium, and slow, corresponding to half-lives on the order of seconds, a few minutes, and several minutes, respectively~\cite{washout-tomitani, mizuno-brain-thigh}. The fast component is thought to reflect the immediate outflow of positron emitters generated within large blood vessels, while the medium and slow components are associated with diffusion and microcirculation, eventually carrying positron emitters to larger blood vessels and out of the tissue, and cellular metabolism~\cite{washout-vascular}. Biological washout occurs simultaneously with the intrinsic, physical decay of these radioactive isotopes. The effect of biological washout on the observed activity over time is illustrated in Figure~\ref{fig:washout-decay}.

While biological washout is often regarded as a limitation for PET-based treatment verification~\cite{washout-limitation}, it has also been recognized as a potential surrogate biomarker for tumor characteristics such as metabolic activity and vascular status~\cite{parodi2015Vision, kira-15o, 11c-perfusion}. Specifically, washout dynamics reveal the metabolic response of treated tissues~\cite{fiedler2008-12C, nishio2010-proton} and effectively reflect vascularity in both perfused and necrotic tumor regions~\cite{washout-vascular, nishio2010-proton, washout-vascular-mri}.

Previous studies on biological washout in proton therapy or hadron therapy mainly considered tumor-wide assessments due to the low PET count statistics of these acquisitions, which limit reliable washout parameter estimates~\cite{kira-15o, 11c-perfusion, washout-vascular-mri}. However, intratumoral heterogeneity in metabolism, microenvironment, and vascular status, previously assessed with PET/CT and PET/MRI imaging, has shown prognostic value for treatment response, resistance, and overall survival~\cite{heterogeneity-microenvironment, heterogeneity-imaging, heterogeneity-uptake, heterogeneity-pet-review, heterogeneity-resistance, heterogeneity-partitioning, heterogeneity-clustering, heterogeneity-subregional, heterogeneity-signatures, heterogeneity-functional, heterogeneity-radiogenomics, heterogeneity-pet-mr-breast}. 
Intratumoral heterogeneity offers additional information beyond traditional treatment response biomarkers (e.g., tumor size), enabling improved risk stratification and the non-invasive detection of aggressive or radioresistant (e.g., hypoxic) tumor regions. Furthermore, improving the spatial characterization of washout kinetics can enhance PET-based dose verification \cite{prototwin-pet}.

\begin{figure*}[h]
    \centering
    \includegraphics[width=\linewidth]{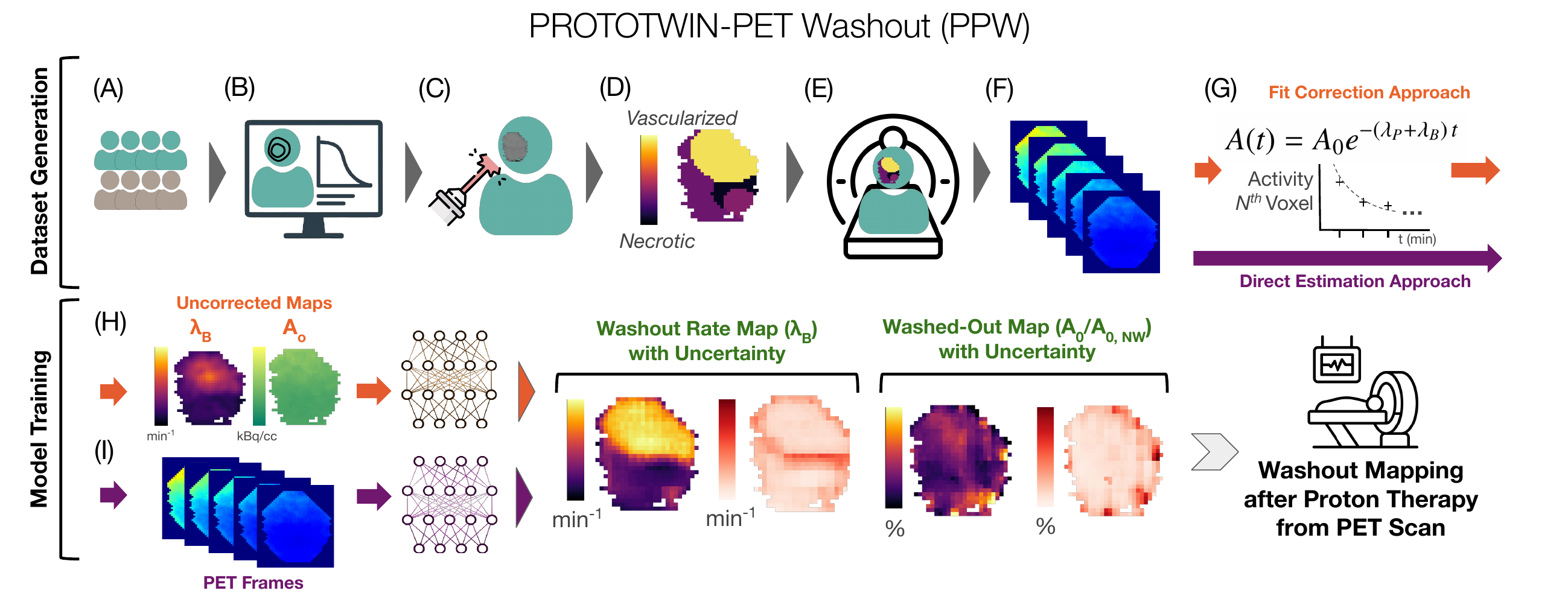}
    \caption{Schematic overview of the PROTOTWIN-PET Washout framework, encompassing dataset generation (top row) and model training (bottom row) for generating voxel-wise biological washout rate and washed-out maps with uncertainty quantification. Dataset generation comprises: (A) patient selection  (four male, four female head-and-neck cancer patients), (B) treatment planning, (C) Monte Carlo simulation of proton therapy (D) introduction of intratumoral washout heterogeneities, creating 75 digital twins per patient and defining ground truths maps, (E) positron emission tomography (PET) acquisition simulation (30-minute scan, 15 minutes post-treatment), and (F) dynamic PET reconstruction using five 6-minute frames. Two map estimation approaches are compared: (G) fit correction (orange), which involves voxel-wise fitting of the activity decay curve $A(t)$, where $\lambda_B$ and $\lambda_P$ denote the slow biological washout rate and the physical decay rate of $^{11}\text{C}$, respectively, followed by (H) model training to correct the fitted parameter map; and (I) direct estimation, where the models learn to estimate the washout rate and washed-out maps directly from the PET frames obtained in step (F). Both approaches provide uncertainty maps (red colormap), using a loss function combining variance-weighted negative log-likelihood ($\beta$-NLL) and structural similarity index metric (SSIM) terms. To generate uncorrected washed-out maps ($A_0/A_{0, \text{NW}}$) for the fit correction approach, steps (D) through (G) are completed to obtain the uncorrected $A_0$ map. Then, to obtain $A_{0, \text{NW}}$, which is the hypothetical activity distribution that would be observed at the start of imaging in the absence of any washout, steps (E) through (G) are repeated for the same tumor sample, skipping step (D).}
    \label{fig:framework}
\end{figure*}

To address the challenge of low SNR in post-proton therapy PET imaging, which limits the reliable quantification and mapping of biological washout parameters, we introduce the PROTOTWIN-PET Washout (PPW) framework (see Figure \ref{fig:framework}). Using Monte Carlo simulations of proton therapy and offline PET imaging in head-and-neck cancer patients with synthetically induced intratumoral heterogeneities, we train an uncertainty-aware deep learning model to accurately map biological washout kinetics. This framework leverages the increased sensitivity of modern time-of-flight (TOF) PET scanners along with the denoising and resolution recovery capabilities of deep learning~\cite{deep-learning-denoising}, allowing for the non-invasive assessment of tumor status, treatment response, and intratumoral heterogeneity without additional radiotracers.



\section{Methods}

This section details the dataset generation, model training and inference, validation strategy, and performance metrics employed in the PROTOTWIN-PET Washout (PPW) framework. The models are trained to estimate washout parameter maps from PET imaging after proton therapy.


\subsection{Dataset Generation}
The dataset used to train the deep learning model is generated through the following steps, illustrated in the top row of Figure~\ref{fig:framework} and detailed below.
\begin{enumerate}[noitemsep, topsep=0pt]
    \item \textbf{Patient selection:} A dataset comprising CT scans with expert-contoured radiotherapy structures is compiled for eight head-and-neck cancer patients (four male, four female), sourced from the TCIA Head-Neck-PET-CT dataset~\cite{head-dataset} (Figure~\ref{fig:framework}A).  Although sex-based differences in biological washout dynamics are not expected to be significant, an equal representation of male and female patients was included to ensure a balanced representation of anatomical variability during model training. 
    
    To expose the model to a range of tumor sizes and stages, the eight patients were selected such that two patients were included for each T stage: T1, T2, T3, and T4, where the T stage denotes the size and local extent of the primary tumor~\cite{tumor-staging}. Adding further patients did not measurably improve model performance, whereas reducing their number resulted in decreased performance. Head-and-neck patients were chosen because they are among the most frequently treated in proton therapy~\cite{head-and-neck-pt}. 
    
    \item \textbf{Treatment planning:} For each patient, proton therapy treatment plans are optimized based on their CT scans and radiotherapy structure segmentations with the open-source treatment planning toolkit MatRad~\cite{matrad} (Figure~\ref{fig:framework}B). Each treatment consists of two anterior-oblique fields and delivers a dose of 2.18 Gy.
    
    \item \textbf{Proton therapy simulation:} Using the optimized treatment plans, proton therapy is simulated with FRED~\cite{fred-activation}, a GPU-based Monte Carlo simulator for proton therapy, to generate a map of produced positron emitters  (Figure~\ref{fig:framework}C). Each treatment considers irradiation and field setup times of 2 minutes, following common clinical practice~\cite{head-multifield, treatment-process-time}.
    
    \item \textbf{Introducing intratumoral heterogeneities:}
    Intratumoral heterogeneities will be identified with PET based on spatiotemporal differences in measured activity (Figure~\ref{fig:framework}D). To induce such heterogeneity, different biological washout parameters are applied across distinct tumor regions. As described in~\cite{mizuno-brain-thigh} and illustrated in Figure \ref{fig:washout-decay}, the induced activity following proton therapy $A(t)$ can be expressed as:

    \begin{equation}
    \begin{aligned}
    A(t) = \sum_{i=1}^{N} A_{0, i} e^{-\lambda_{P,i} \,t} \bigl( &
    M_{s,i}\, e^{-\lambda_{B,s, i}\,t} +  \\
    M_{m,i}\, e^{-\lambda_{B,m, i}\,t}
    & + M_{f,i}\, e^{-\lambda_{B,f, i}\,t} \bigr)
    \end{aligned}
    \label{eq:total-activity}
    \end{equation}

    
   Here, the index \( i \) denotes summation over the \( N=4 \) considered positron emitters, specifically \( ^{11}\text{C} \), \( ^{13}\text{N} \), \( ^{15}\text{O} \), and \( ^{38}\text{K} \). Collectively, they constitute over 99\% of the produced isotopes. For each positron emitter \( i \):
   
    - \( A_{0, i} \) represents the initial activity at the start of the imaging period (\( t = 0 \)). 
    
    - \( \lambda_{P,i} \) is the physical decay constant, with values \(0.0492\,\text{min}^{-1}\) for \(^{11}\text{C}\),
\(0.1003 \,\text{min}^{-1}\) for \(^{13}\text{N}\), \(0.4907\,\text{min}^{-1}\) for \(^{15}\text{O}\), and
\(0.1308\,\text{min}^{-1}\) for \(^{38}\text{K}\)~\cite{phys-decay}.

    - \( \lambda_{B,s,i} \), \( \lambda_{B,m,i} \), and \( \lambda_{B,f,i} \) represent the slow, medium, and fast biological washout rates, respectively.
    
    - \( M_{s,i} \), \( M_{m,i} \), and \( M_{f,i} \) indicate the fractions of positron emitters undergoing slow, medium, and fast biological washout, respectively, at the start of imaging. Defining the fractions at imaging onset ensures that \( A_{0, i} \) represents the activity at \( t = 0 \). The sum of these three fractions is equal to one.

    Based on the relative abundances of positron emitters produced during treatment~\cite{prototwin-pet}, adopting an offline imaging approach and starting the PET acquisition 15 minutes post-treatment results in approximately 80\% of the remaining activity originating from $^{11}\text{C}$.  Furthermore, a 15-minute delay provides enough time to transfer the patient from the proton therapy room to the PET scanner, facilitating integration of the framework into clinical practice. 
    
    With this delay, the fast and medium biological washout components of $^{11}\text{C}$ have largely decayed at the start of imaging \sloppy(\(M_{f,^{11}\text{C}},~M_{m,^{11}\text{C}}~\ll~M_{s,^{11}\text{C}}\)), enabling the analysis to focus on the slow washout component, characterized by the rate constant $\lambda_{B,s,^{11}\text{C}}$, which dominates the observed activity signal~\cite{bauer-11c-brain} (see~Figure~\ref{fig:washout-decay}). 
    This reduces uncertainty in the simulated positron emitter map, as only the production cross-section of $^{11}\text{C}$ isotopes remains relevant~\cite{janis-proton-cross-sections}. Consequently, Equation \ref{eq:total-activity} simplifies to the form presented in Equation \ref{eq:offline-activity}. For conciseness, we will denote $\lambda_{P,^{11}\text{C}}$ as $\lambda_{P}$; $A_{0, ^{11}\text{C}}$ as $A_0$; and $\lambda_{B, s, ^{11}\text{C}}$ as $\lambda_{B}$ hereafter.
    \begin{equation}
    A(t) = A_0 e^{-(\lambda_{P} + \lambda_{B})\,t}
    \label{eq:offline-activity}
    \end{equation}

    For each tumor, one to five regions with varying values of $A_0$ and $\lambda_B$ are defined to generate ground truth maps. The assigned values for $\lambda_{B}$ range from 0.0 $\text{min}^{-1}$ (indicative of no washout, as in necrotic regions) to $45 \times 10^{-3} \text{min}^{-1}$ for highly vascularized regions. The upper limit was set slightly higher than previously reported human values, which reached approximately $35 \times 10^{-3} \text{min}^{-1}$~\cite{bauer-11c-brain}. This is because earlier studies reported values averaged across tissues, whereas localized subregions within tumors may exhibit higher washout rates. Moreover, a higher upper limit can account for residual contributions from other isotopes. 
        
    The initial activity distribution at the start of imaging, $A_0$, is derived from the spatial map of produced positron emitters obtained in the preceding step, accounting for the reduction in activity between the end of treatment and the onset of imaging due to both physical and biological clearance. To introduce variations in $A_0$ across tumoral regions, beyond those arising from slow-washout differences, we vary washout rates and component fractions of the fast and medium washout components (see Equation \ref{eq:total-activity}) within $\pm$50\% of their default values. The default values are based on values reported in previous studies~\cite{washout-values-1, washout-values-2}.
    
    Slow, medium, and fast biological washout processes differ in nature and may offer different information about tumor status. Therefore, we aim to analyze them in the following manner:
    
    \begin{description}[leftmargin=0pt, labelsep=1em, itemsep=0.5em, parsep=0pt, topsep=1.0em, partopsep=0pt]
        \item[Washout rate maps (slow components):]  Washout rate maps represent the voxel-wise quantification of the slow washout rate $\lambda_{B}$ (see Equation~\ref{eq:offline-activity}), reflecting the rate at which isotopes are cleared during imaging. As slow washout rates have been linked to tumor status~\cite{washout-vascular}, they constitute the primary focus of this study.
        
        \item[Washed-out maps (medium and fast components):]    
        Faster washout components (specifically the medium and fast components, as defined in Equation~\ref{eq:total-activity}) may offer additional insights, but extracting the specific washout rates ($\lambda_B$) and component fractions ($M$) for the mixture of positron emitters and components in Equation~\ref{eq:total-activity} from the dynamic PET data is unfeasible, particularly with an offline approach. Nevertheless, these factors influence the magnitude of the initial activity measured at the start of imaging ($A_0$ in Equation \ref{eq:offline-activity}).
        
        More specifically, we define the \textit{washed-out} maps as the voxel-wise quantification of the ratio $A_0 / A_{0, \text{NW}}$, where $A_0$ is the observed initial activity distribution and $A_{0, \text{NW}}$  represents the hypothetical activity distribution in the absence of biological washout (with ``NW" denoting no washout). Therefore, the washed-out maps represent the fraction of the activity that has not been washed out at the onset of imaging (see Figure \ref{fig:washout-decay}). In these maps, lower values indicate a greater influence of medium and fast biological washout components, meaning more positron emitters have already been washed out ($A_0$ is significantly lower than $A_{0, \text{NW}}$), whereas higher values indicate lower influence of fast and medium washout components ($A_0$ remains closer to $A_{0, \text{NW}}$).
        
        The distribution $A_{0, \text{NW}}$ is computed analogously to $A_0$, based on the spatial map of produced positron emitters, but accounting solely for physical decay occurring between the end of treatment and the start of imaging, excluding any biological washout effects.
        
        We introduce this metric as a potential proxy for the washout effect of medium and fast components of $^{11}\text{C}$ washout. While it has not been previously defined or studied in relation to tumoral status, we hypothesize it as an indicator of washout dynamics.
    \end{description}
    
    The introduced regions have varied, irregular spheroid shapes~\cite{spheroid_repo}, resembling those observed in prior studies~\cite{heterogeneity-subregional, heterogeneity-functional, heterogeneity-mri}. For each patient, 75 heterogeneous tumor instances are generated, serving as their digital twins and providing ground truths. With eight patients, this yields a total of 600 training cases. Increasing the number of digital twins did not lead to further performance gains, whereas reducing their number resulted in decreased performance.

    \item \textbf{PET acquisition simulation:} 
    30-minute TOF-PET acquisitions were simulated on the Siemens Biograph Vision PET/CT scanner (hereafter referred to as Vision), starting 15 minutes after treatment
     (Figure~\ref{fig:framework}E). The framework models crystal size, detector resolution, and depth of interaction (DOI) effects.
     
     Using these parameters and the ground truth maps for $A_0$, $A_{0,\text{NW}}$, and $\lambda_B$, defined in the previous step for each digital twin, the GPU-based MCGPU-PET simulator~\cite{mcgpu-pet} is used to simulate the PET acquisitions and generate corresponding lists of detected coincidence events. On average, each acquisition resulted in approximately 12 million coincidence events.
    
    \item \textbf{PET data processing:} 

     Five 6-minute PET frames were reconstructed from the list-mode PET data using the GPU-based software \textit{parallelproj}~\cite{parallelproj} (Figure~\ref{fig:framework}F). This frame length was found to provide an optimal balance between capturing sufficient time points for robust fitting and minimizing noise by avoiding excessively short frames~\cite{imaging-time-points}. 
     
     Due to the low PET count statistics and high noise, a single iteration of the MLEM algorithm was found to be optimal for reconstruction. The PET image reconstruction incorporates both the point spread function (PSF) and time-of-flight (TOF) information.
     
     The PET frames, washed-out maps ($A_0 / A_{0, \text{NW}}$), and washout rate maps ($\lambda_B$) have the same resolution as the CT scans, with a voxel size of $1.95~\times~1.95~\times~1.5$~mm, and dimensions of $64~\times~64~\times~64$ voxels, which was sufficient to fully enclose the tumor in all cases. To isolate the region of interest, all voxels beyond a four-voxel margin surrounding the tumor are set to zero. The four-voxel margin is chosen to ensure inclusion of any potential information extending beyond the tumor boundaries due to noise and spillover effects from the image PSF. 

\end{enumerate}

At this stage, a dataset comprising realistic post-proton therapy PET frames and corresponding ground truth washout rate and washed-out maps has been generated. This enables the training of a deep learning model to estimate washout rate and washed-out maps from the PET data. 

\subsection{Model Training}

\subsubsection{Map Estimation Approach}
We evaluate two approaches for training the deep learning model: (1)~using uncorrected washed-out and washout rate maps as input, referred to as the \textit{Fit Correction} approach, and (2) using PET frames directly, referred to as the \textit{Direct Estimation} approach.

\begin{description}[leftmargin=0pt, labelsep=1em, itemsep=0.5em, parsep=0pt, topsep=1em, partopsep=0pt]
    \item[Fit Correction:] For each voxel, the exponential decay model of Equation \ref{eq:offline-activity} is fitted to the measured PET intensities across all frames  (Figure~\ref{fig:framework}G). From this fit, the slow washout rate $\lambda_B$ and the initial activity $A_0$ are estimated and mapped  (Figure~\ref{fig:framework}H). For the washed-out maps ($A_0/A_{0, \text{NW}}$), the decay model is also fitted to the acquisition where no washout is considered, yielding $A_{0, \text{NW}}$ maps.
    
    Time–activity curves were fitted for each voxel by nonlinear least‐squares regression using the \texttt{curve\_fit} function from the \texttt{scipy.optimize} module of the SciPy library~\cite{scipy}. For voxels with insufficient PET counts to reliably fit the decay curve, defined as voxels where the uncertainty of the fitted parameters exceeds 20\%, values are computed by averaging neighboring voxels within a $3~\times~3~\times~3$ kernel. 
    
    Although these fitted maps already represent washout rate and washed-out maps, the SNR is low and they suffer from partial volume effects commonly encountered in PET imaging~\cite{partial-volume}. Specifically, the PSF of the images causes PET counts to spill over into neighboring regions, leading to biased values. Therefore, we call these uncorrected maps. 
    
    A deep learning model is then trained to correct these estimates, reducing noise and bias. The model estimates $\lambda_B$ (washout rate) and $A_{0} / A_{0, \text{NW}}$ (washed-out) maps, functioning similarly to a deep learning-based denoising and deblurring tool. 
    
    \item[Direct Estimation (from PET frames):]
    In this approach, the deep learning model is trained to estimate the washout rate maps directly from the PET image frames, bypassing the need for explicit curve fitting  (Figure~\ref{fig:framework}I). This method is only tested on the washout rate maps, as the task is less complex than that of the washed-out maps, which also requires PET frames without washout.
    
    A key limitation of this method is its reduced generalizability, as it requires a fixed number of frames with a specific duration, making it less flexible across different acquisition protocols. Additionally, this approach presents a greater challenge for the model, as it lacks the implicit prior knowledge provided by activity curve fitting. Instead, the model must learn to extract the washout rate maps directly from the raw PET frames.
    
    However, this method offers a significant advantage: it eliminates the need to fit decay curves for each voxel, reducing computational overhead. Moreover, it retains the full spatiotemporal information from the PET reconstruction, potentially capturing features that might be lost in curve fitting.
\end{description}

\subsubsection{Uncertainty Quantification and Loss Function}

Trustworthiness is a critical concern in deep learning, particularly for medical applications~\cite{uncertainty-clinical}. One approach to improve trust in a model's predictions is to quantify and report associated uncertainty estimates. To this end, the models provide both epistemic and aleatoric uncertainties, as described below.

\textbf{Epistemic uncertainty} is quantified using Monte Carlo (MC) dropout with a dropout rate of 0.2 and 20 iterations~\cite{mc-dropout}. This means that the model is run 20 times on the same input, with 20\% of the neurons randomly deactivated in each run. The final estimate is computed as the mean of these 20 outputs, while the epistemic uncertainty is defined as the standard deviation among them ($\hat{\sigma}_{\text{epistemic}}$). High epistemic uncertainty reflects model instability and lack of knowledge and can be reduced by increasing training data or improving the model. Both the dropout rate and the number of iterations were experimentally optimized and are in line with standard values in medical imaging applications~\cite{mc-dropout}.

\textbf{Aleatoric uncertainty} reflects the intrinsic uncertainty due to the noise in the PET data~\cite{aleatoric} and is estimated directly by the model, which outputs both the washed-out or washout rate map and the corresponding aleatoric variance (see Figure~\ref{fig:architecture}). These outputs are learned jointly by minimizing the variance-weighted negative log-likelihood (NLL) loss, denoted as $\beta$-NLL~\cite{beta-nll}. 

\begin{equation}
  \beta\text{-NLL}=
       \lfloor\hat{\sigma}^{2\beta}_{x}\rfloor
       \,\frac{1}{2}\,
       \left[
         \frac{(y-x)^{2}}
              {\hat{\sigma}^{2}_{x}}
         +\,
         \ln\left(\hat{\sigma}^{2}_{x}\right)
       \right]
    \label{eq:beta-NLL}
\end{equation}

Here, $x$ denotes the estimated map, $y$ the ground truth map, $\hat{\sigma}^2_{x}$ the predicted aleatoric variance, and $\hat{\sigma}_{x} \coloneqq \hat{\sigma}_{\text{aleatoric}}$ the aleatoric uncertainty. 

The $\beta$-NLL loss contains two additive terms: a residual term inversely weighted by the predicted variance, which penalizes confident but inaccurate estimations; and a regularization term preventing the model from simply assigning high uncertainty to all estimations to minimize the first term. The loss is scaled by the detached factor $\lfloor\hat{\sigma}^{2\beta}_{x}\rfloor$, which controls the sensitivity to variance estimation errors. The \textit{stop gradient} operation $\lfloor\cdot\rfloor$ ensures that gradients do not flow through the $\hat{\sigma}^2_{x,i}$ term, preventing it from being driven to degenerate extremes (either $0$ or $\infty$). The hyperparameter $\beta$ is set to 0.25, as determined empirically and based on its effectiveness in the original study~\cite{beta-nll}

Finally, the total predictive uncertainty $\hat{\sigma}_{\text{total}}$ is obtained by combining aleatoric and epistemic components via quadratic summation, under the assumption of independence~\cite{aleatoric}. This uncertainty formulation ensures that both data-dependent (aleatoric) and model-related (epistemic) sources of uncertainty contribute to the final uncertainty estimate.

\begin{equation}
\hat{\sigma}_{\text{total}} = \sqrt{\hat{\sigma}^2_{\text{aleatoric}} + \hat{\sigma}^2_{\text{epistemic}}}
\label{eq:total_uncertainty}
\end{equation}

The total loss function is a weighted sum of two loss functions. The first loss function is $\beta$-NLL (see Equation \ref{eq:beta-NLL}), while the second is the structural similarity index measure (SSIM):

\begin{equation}
  \operatorname{SSIM}(x,y)=
  \frac{(2\mu_x\mu_y + C_1)(2\sigma_{xy} + C_2)}
       {(\mu_x^{2} + \mu_y^{2} + C_1)(\sigma_x^{2} + \sigma_y^{2} + C_2)} ,
\label{eq:ssim}
\end{equation}
where $x$ is the estimated map and $y$ is the ground truth map. The \(\mu_x,\mu_y\) are local means, \(\sigma_x^{2},\sigma_y^{2}\) local variances,
and \(\sigma_{xy}\) the local covariance, all evaluated over the same sliding \(11~\times~11~\times~11\) Gaussian window.  $C_1$ and $C_2$ are stabilizing constants preventing division by zero when local variance is close to zero.

The SSIM enhances the visual correspondence between the estimated and ground truth maps, and ranges from $-$1 to 1, where higher values indicate more similar structure, luminance, and contrast~\cite{ssim}. The optimal weighting for the SSIM loss is empirically determined to be 0.1, while the $\beta$-NLL loss is assigned a weight of 0.9.

\begin{equation}
    \mathcal{L}_{\text{total}} = 0.1 \cdot \text{SSIM} + 0.9 \cdot \beta\text{-NLL}
    \label{eq:total-loss}
\end{equation}


\subsubsection{Model Architecture and Training Parameters}
\begin{figure}[h]
    \centering
\includegraphics[width=\linewidth]{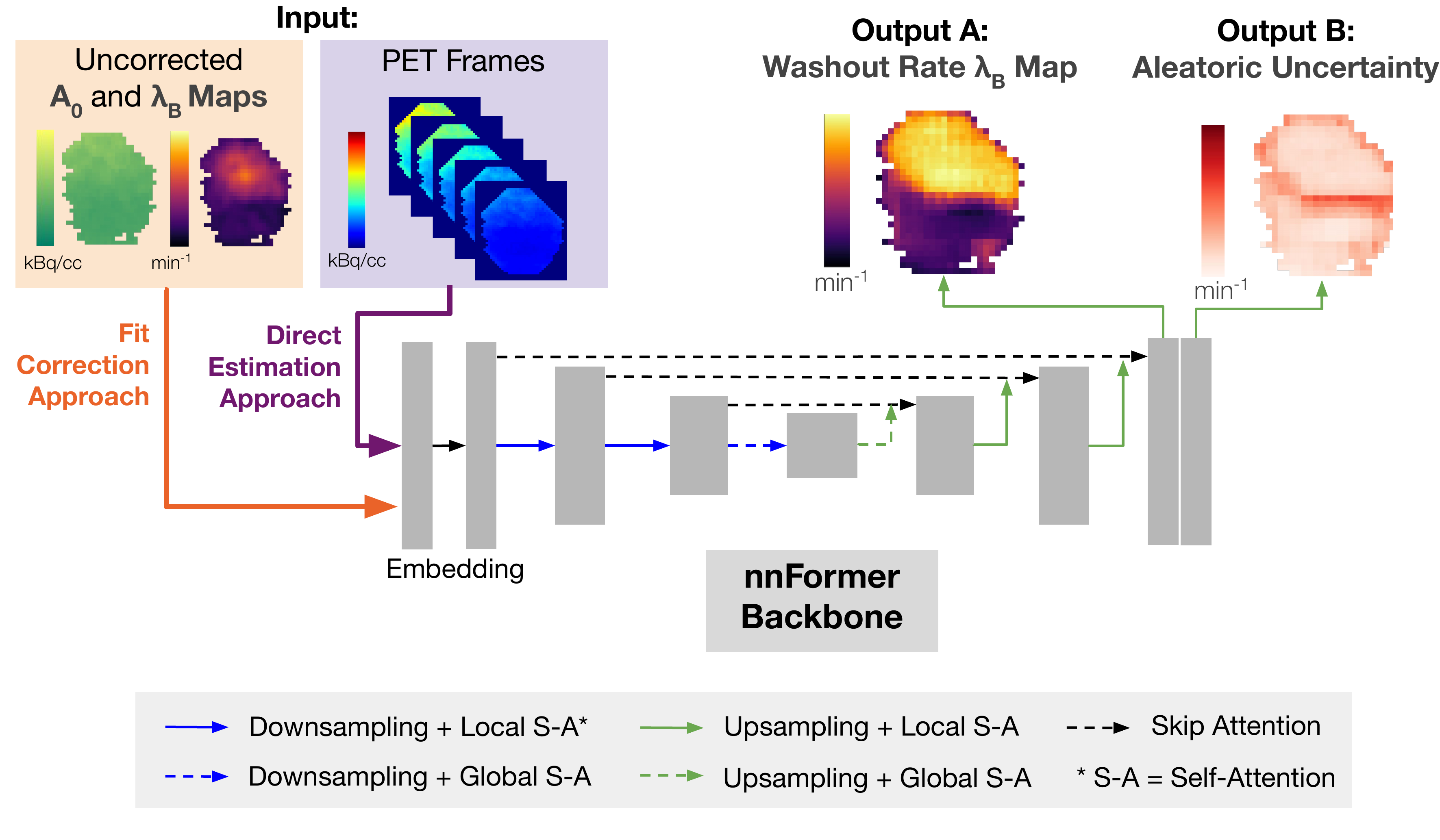}
    \caption{Architecture of the nnFormer model, designed to output both the washout rate map ($\lambda_B$) and its aleatoric uncertainty. The model inputs are either the uncorrected maps (fit correction approach) or the PET frames (direct estimation approach). These inputs undergo processing through convolutional and self-attention layers, compressing and expanding spatial and channel dimensions to extract diverse and rich features. All the details regarding the model architecture can be referenced in the original publication~\cite{nnformer}. For simplicity, only the estimation of the washout rate ($\lambda_B$) map is shown. The architecture used for estimating the washed-out map ($A_0 / A_{0, \text{NW}}$) is identical, except it is trained only using the fit correction approach.}
    \label{fig:architecture}
\end{figure}

The selected neural network architecture, used for both approaches and which consistently yielded the best results, is based on the nnFormer model, which incorporates self-attention mechanisms within a UNet backbone~\cite{nnformer} (see Figure~\ref{fig:architecture}). The model is trained using the AdamW optimizer~\cite{adamw}, with the initial learning rate determined experimentally and assigned a value of 0.0001. Training extends for up to 1000 and 2500 epochs for washout rate and washed-out maps, respectively. Early stopping is triggered if the loss fails to improve for 150 epochs.  A cosine annealing scheduler adjusts the learning rate, with a complete period covering all epochs and a minimum learning rate set to 20\% of the initial learning rate value~\cite{cosine-scheduler}. 

Input maps undergo min-max scaling~\cite{minmax}, with scaling parameters defined by the minimum and maximum decay rates for the washout rate maps, and by the range [0, 100]\% for the washed-out maps. Data augmentation techniques from the MONAI transforms library~\cite{monai} were applied to enhance model generalizability across diverse tumor types and morphologies. These included rotations (probability = 50\%, angle $\leq$ 10$^\circ$), flipping along all three axes (probability = 30\%), and elastic deformations (probability = 20\%, $\sigma \in [6,10]$, magnitude $\in [100,150]$).

\subsection{Model Inference}

Once the models are trained, they can be implemented for washout rate and washed-out mapping after proton therapy. A PET scan should be performed, and the PET frames reconstructed from the list-mode data. If the direct estimation approach were implemented, these frames could be directly provided to the model to estimate washout rate maps. If the fit correction were implemented, a voxel-wise activity fit of Equation \ref{eq:offline-activity} should be performed to generate model inputs (uncorrected $\lambda_B$ and $A_0$ maps). When applying the fit correction approach to washed-out maps, the initial activity without washout ($A_{0, \text{NW}}$) should be estimated based on implementation parameters, incorporating a patient-specific PET and proton therapy simulation and considering that there is no biological washout (i.e. following steps (C), (E), (F), (G) in Figure~\ref{fig:framework}).

The framework outputs visualizations of uncorrected (if implementing the fit correction approach) and estimated washout rate and washed-out maps. The washout rate maps are also thresholded, with voxels discretized and classified into one of three types. These vascularity categories are uniformly distributed between the lower and upper limits of the slow washout rates used during training, ranging from 0.0 $\text{min}^{-1}$ to $45 \times 10^{-3} \text{min}^{-1}$. This discretization produces three categories representing low, medium, and high vascularity, facilitating a more intuitive interpretation of the washout rates within tumor regions and serving as a more interpretable surrogate for intratumoral heterogeneity and vascularity.

\subsection{Model Validation}

The proposed framework was validated on four additional head-and-neck cancer patients (two male, two female), each representing different tumor stages from T1 to T4. 75 digital twins are generated for each patient, resulting in a total of 300 training cases. The model was further evaluated on a single liver cancer patient from the CORT dataset~\cite{cort-dataset}, allowing assessment of its generalizability to a different anatomical region despite being trained on head-and-neck cancer cases.


\subsection{Performance Metrics and Statistical Analysis}

To evaluate the correspondence between estimated and ground truth maps, two primary metrics are used: absolute error (AE) and SSIM. Results are reported as the median values (medAE and medSSIM) along with the interquartile range (IQR), defined between the first (Q$_1$) and third (Q$_3$) quartiles. Both AE and SSIM are only assessed inside the tumor, which is the region of interest and where heterogeneity has been introduced.

The medAE is defined as:
\begin{equation}
\text{medAE}(x, y) = \text{median}\left(|x_j - y_j|\right)
\label{medAE}
\end{equation}

where $j$ represents each map voxel in the test set, $x$ is the estimated map, and $y$ is the ground truth map. medAE and IQR are appropriate due to the left-skewed (Shapiro-Wilks test, $p \ll 0.001$) and zero-bounded distribution of washout rate and washed-out absolute errors (AE). Lower medAE values indicate more accurate estimates. 

The medSSIM quantifies structural similarity between the estimated and ground truth maps (see Equation~\ref{eq:ssim}). medSSIM and IQR are appropriate because the SSIM distributions for washout rate and washed-out maps are right-skewed (Shapiro-Wilks test, $p \ll 0.001$) and bounded above by one. Higher medSSIM values indicate improved similarity. Because SSIM is computed per tumor sample, medSSIM is determined by calculating the median of the SSIM values obtained for each tumor sample in the test set.

Box plots are used to illustrate how AE distributions vary across tumor regions of increasing size, ranging from 1.7 mL to 32.6 mL. Each bin represents a 3.4 mL range in region volume; for example, the boxplot centered at 5.1 mL includes tumors with sizes between 3.4 mL and 6.8 mL, and so forth. The central horizontal line in each box represents the median; the box spans from the first to the third quartile, and the whiskers indicate the range of the errors. The statistical significance of differences between the AE distributions of uncorrected and corrected maps for each region size is evaluated using the two-tailed Wilcoxon signed-rank test~\cite{wilcoxon}.

To assess the uncertainty quantification, a sparsification error plot illustrates how the medAE changes as voxels with higher estimated uncertainty are progressively removed. A decreasing trend in medAE would indicate that the uncertainty estimates effectively identify regions of higher model error~\cite{sparsification}.

\section{Results}

\subsection{Washout Estimation Performance and Approach Comparison}

Table \ref{tab:slow_approach_comparison} presents the performance of the deep learning models for washout rate map estimation, compared against the results obtained by direct fitting to the data. The fit correction model provides the most accurate estimates, as indicated by the lowest AE and highest SSIM, achieving a 60\% reduction in medAE and a 16\% increase in medSSIM relative to uncorrected maps.

\begin{table}[h]
    \centering
    \caption{Comparison of structural similarity index measure (SSIM) and absolute error (AE) metrics for washout rate maps ($\lambda_B$) obtained via three approaches: direct fitting to noisy PET frames (Uncorrected), deep learning-based correction (Fit Correction), and directly estimated with deep learning from PET frames (Direct Estimation). Results are reported as median [Q$_1$~--~Q$_3$] for both SSIM (medSSIM~[Q$_1$~--~Q$_3$]) and absolute error (medAE~[Q$_1$~--~Q$_3$]). SSIM values range from -1 to 1 (higher is better); AE values are non-negative (lower is better). The best performance for each metric is indicated in bold.}
    \resizebox{\linewidth}{!}{ 
        \begin{tabular}{lcc}
            \toprule
            Approach & medSSIM [Q$_1$~--~Q$_3$]  & \makecell{medAE [Q$_1$~--~Q$_3$]\\$(10^{-3}\, \text{min}^{-1})$} \\
            \midrule
            Uncorrected & 0.64 (0.52 -- 0.73) & 5.8 (2.1 -- 14.8) \\
            Fit Correction  & \textbf{0.80 (0.75 -- 0.83)}  & \textbf{2.3 (1.0 -- 5.1)} \\
            Direct Estimation & 0.78 (0.73 -- 0.82) & 3.1 (1.2 -- 8.2) \\
            \bottomrule
        \end{tabular}
        }
    \label{tab:slow_approach_comparison}
\end{table}

Performance metrics for the washed-out map estimations are presented in Table \ref{tab:fast_approach_comparison}. Given the superior performance and generalizability of the Fit Correction approach compared to Direct Estimation, we focus exclusively on Fit Correction for subsequent analyses. In this case, the improvements after fit correction are less pronounced, as reflected by smaller increases in SSIM and reduced decreases in AE. The fit correction achieves a 28\% reduction in medAE and a 10\% increase in medSSIM relative to uncorrected maps.

\begin{table}[h]
    \centering
    \caption{Comparison of structural similarity index measure (SSIM) and absolute error (AE) metrics for washed-out maps ($A_0/A_{0, \text{NW}}$) obtained via two approaches: direct fitting to noisy PET frames (Uncorrected) versus deep learning-based correction (Fit Correction). Results are reported as median [Q$_1$~--~Q$_3$] for both SSIM (medSSIM~[Q$_1$~--~Q$_3$]) and absolute error (medAE~[Q$_1$~--~Q$_3$]). SSIM values range from -1 to 1 (higher is better); AE values are non-negative (lower is better). The best performance for each metric is indicated in bold.}
    \resizebox{\linewidth}{!}{ 
        \begin{tabular}{lcc}
            \toprule
            Approach & medSSIM [Q$_1$~--~Q$_3$]  & medAE [Q$_1$~--~Q$_3$] [\%]\\
            \midrule
            Uncorrected & 0.58 (0.53 -- 0.67) & 2.9 (1.3 -- 5.1) \\
            Fit Correction  &  \textbf{0.68 (0.64 -- 0.73)}  & \textbf{2.1 (0.9 -- 4.0)} \\
            \bottomrule
        \end{tabular}
    }
    \label{tab:fast_approach_comparison}
\end{table}

\subsection{Error Dependence on Region Volume}

\begin{figure*}[h]
    \centering
    \begin{subfigure}{0.49\linewidth}
        \centering
        \includegraphics[width=\linewidth]{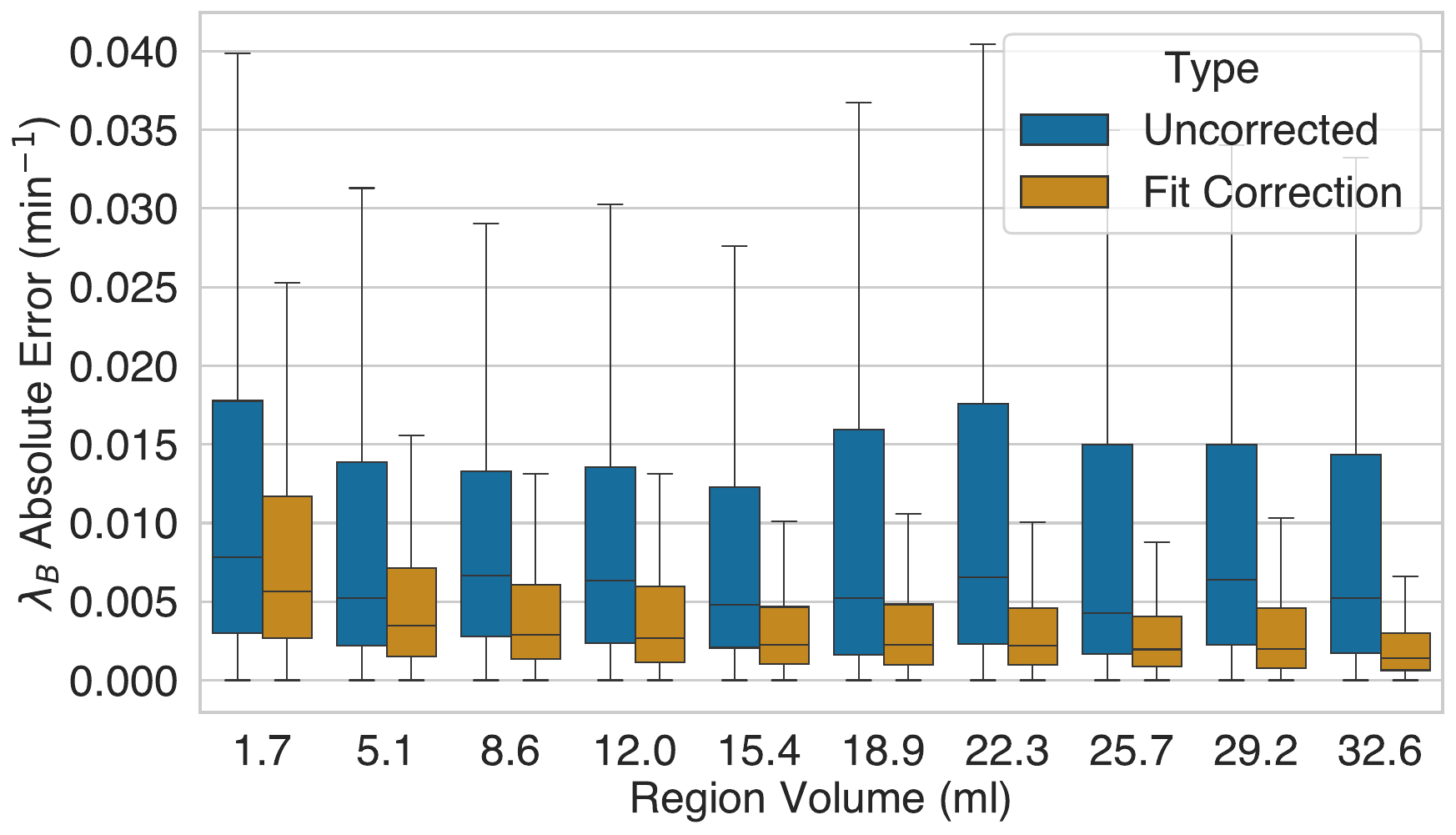}
        \subcaption{Washout rate maps ($\lambda_B$)}
        \label{fig:slow-error-vs-region}
    \end{subfigure}
    \begin{subfigure}{0.49\linewidth}
        \centering
        \includegraphics[width=\linewidth]{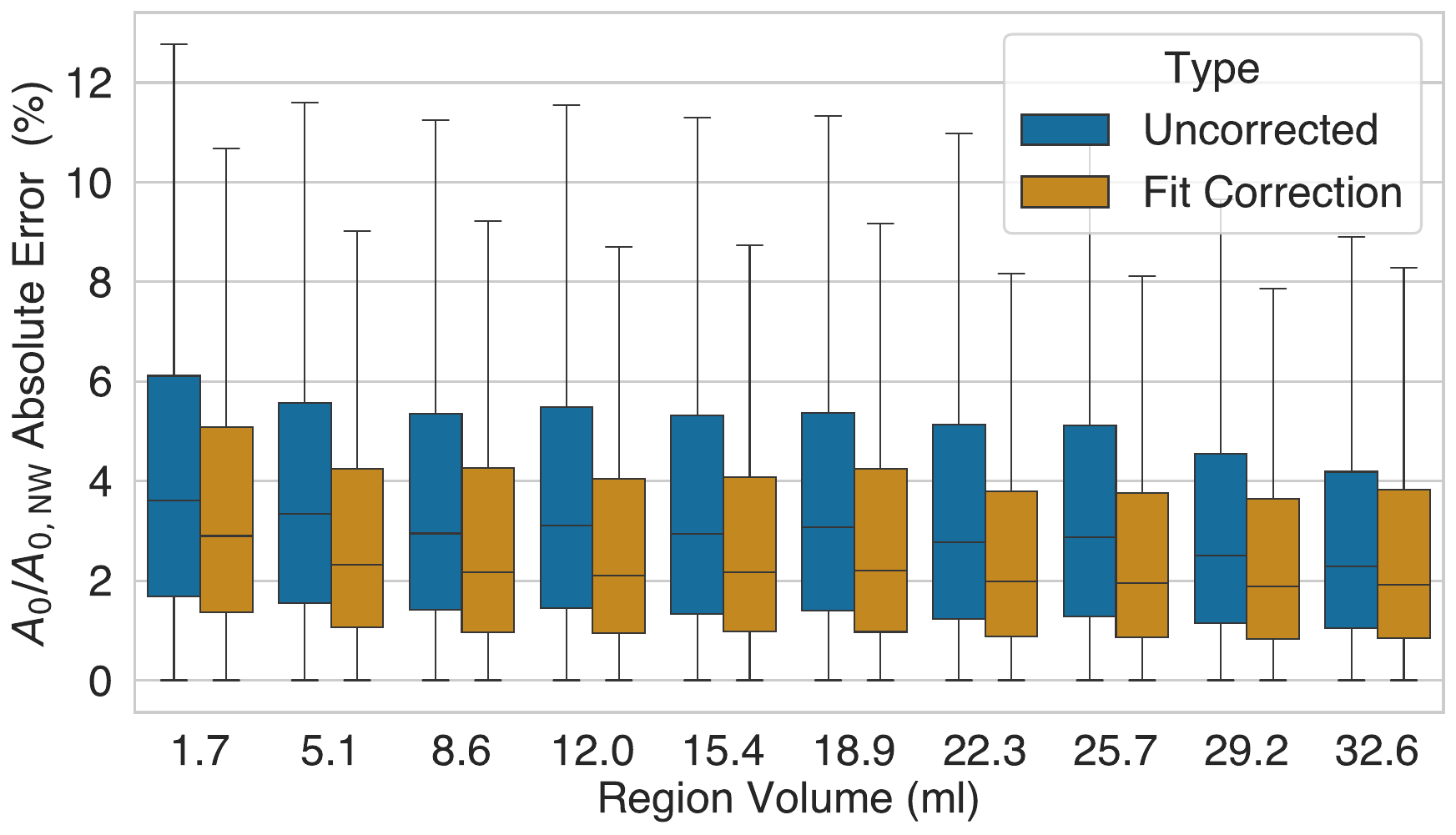}
        \subcaption{Washed-out maps ($A_0/A_{0,\text{NW}}$)}
        \label{fig:fast-error-vs-region}
    \end{subfigure}
    \caption{Absolute error distributions obtained with the fit correction approach and without correction (Uncorrected) across region volumes ranging from 1.7\,mL to 32.6\,mL. For both (a) washout-rate and (b) washed-out maps, corrected errors are significantly lower in every volume bin (two-sided Wilcoxon signed-rank test, $p \ll 0.001$). Voxel counts per bin (ascending region volume) are
    n~=~72,296; 110,092; 100,558; 144,316; 213,747; 165,699; 197,081; 127,788; 227,740; 147,328, with each voxel contributing paired error measurements.}
    \label{fig:error-vs-region}
\end{figure*}

Figure \ref{fig:slow-error-vs-region} presents boxplots of the absolute errors in estimating the washout rate maps, comparing uncorrected and fit correction approaches across regions of increasing volume. As expected, smaller regions exhibit higher errors due to the greater difficulty of extracting signal from noise. Figure \ref{fig:fast-error-vs-region} presents the same analysis for the washed-out map estimates.

\subsection{Visualization of Maps}
\begin{figure*}[h]
    \centering
    \includegraphics[width=\linewidth]{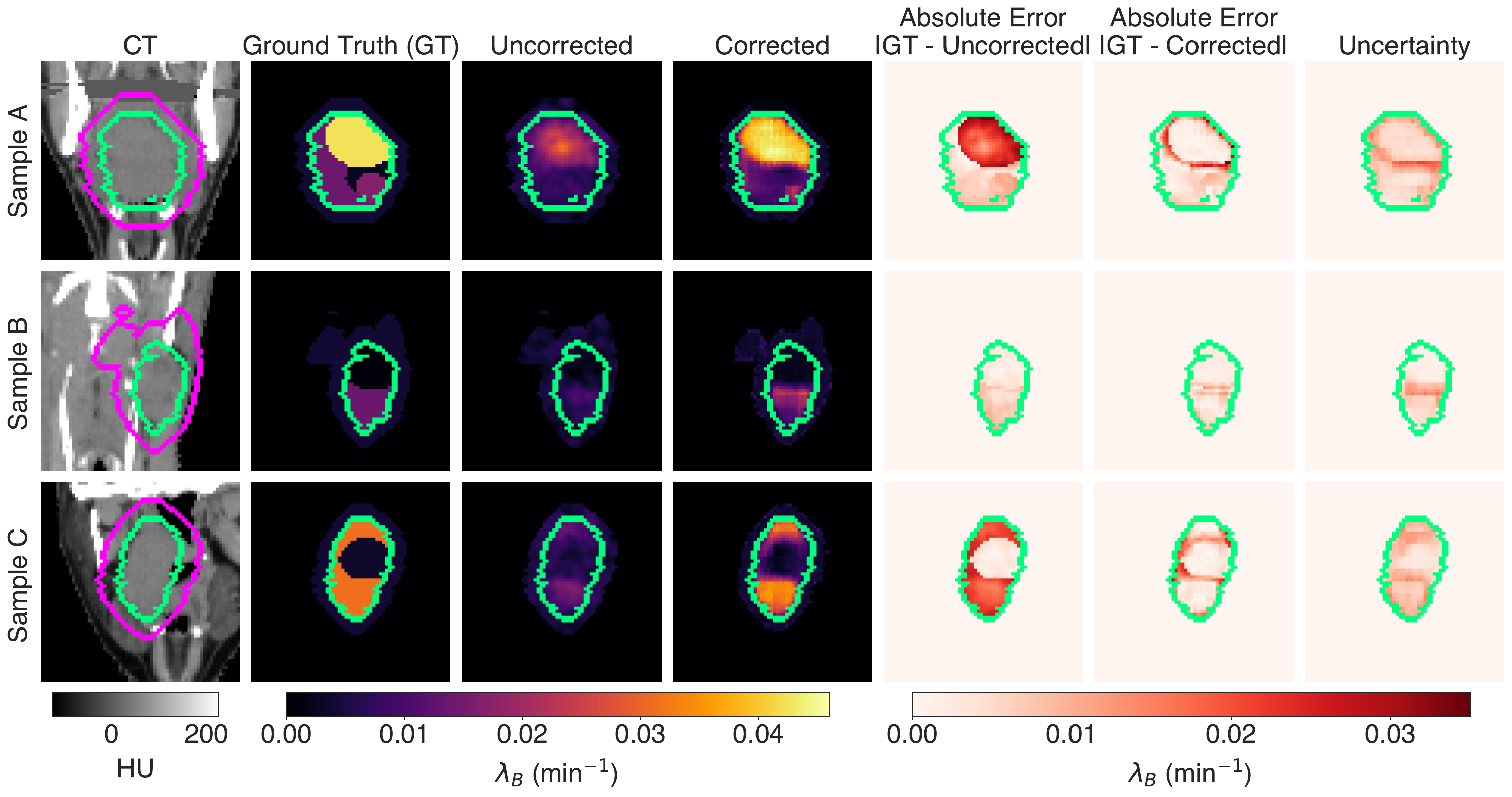}
    \caption{Washout rate maps ($\lambda_B$) obtained with the fit correction approach (Corrected) and without correction (Uncorrected) are compared with the ground truth (GT) for three tumor samples in the test set, shown in a coronal view. The CT scan is included for anatomical context, with the tumor outline (green) and the total region with margins provided to the model (pink) overlaid. For clarity, uncertainty and error maps are displayed using a distinct colormap from the washout rate maps.}
    \label{fig:sample-slow}
\end{figure*}

\begin{figure*}[h]
    \centering
    \includegraphics[width=\linewidth]{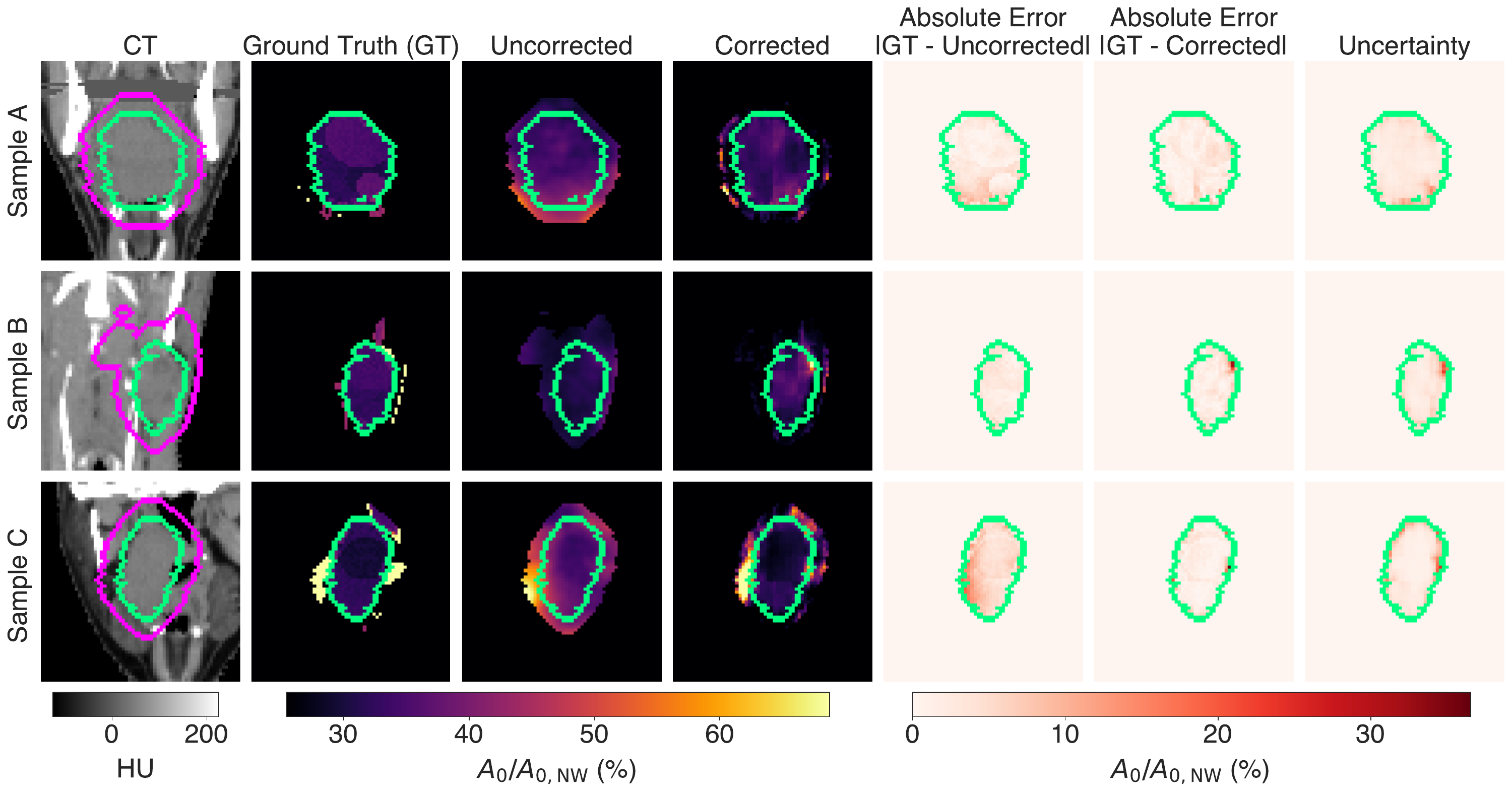}
    \caption{Washed-out maps ($A_0 / A_{0, \text{NW}}$) obtained with the fit correction approach (Corrected) and without correction (Uncorrected) are compared with the ground truth (GT) for three tumor samples in the test set, shown in a coronal view. The CT scan is included for anatomical context, with the tumor outline (green) and the total region with margins provided to the model (pink) overlaid. For clarity, uncertainty and error maps are displayed using a distinct colormap from the washed-out maps.}
    \label{fig:sample-fast}
\end{figure*}

In Figure \ref{fig:sample-slow}, the uncorrected and corrected washout rate maps are compared for three representative tumor samples from the test set, showcasing the improvement in region separation and the alignment between areas of higher uncertainty and higher errors. These areas are mainly located on the edges of the region's boundaries. Figure \ref{fig:sample-fast} presents a similar comparison for the washed-out maps. However, in these cases, the improvements after correction are less pronounced.

\subsection{Uncertainty Quantification}

\begin{figure}[h]
    \centering
    \includegraphics[width=\linewidth]{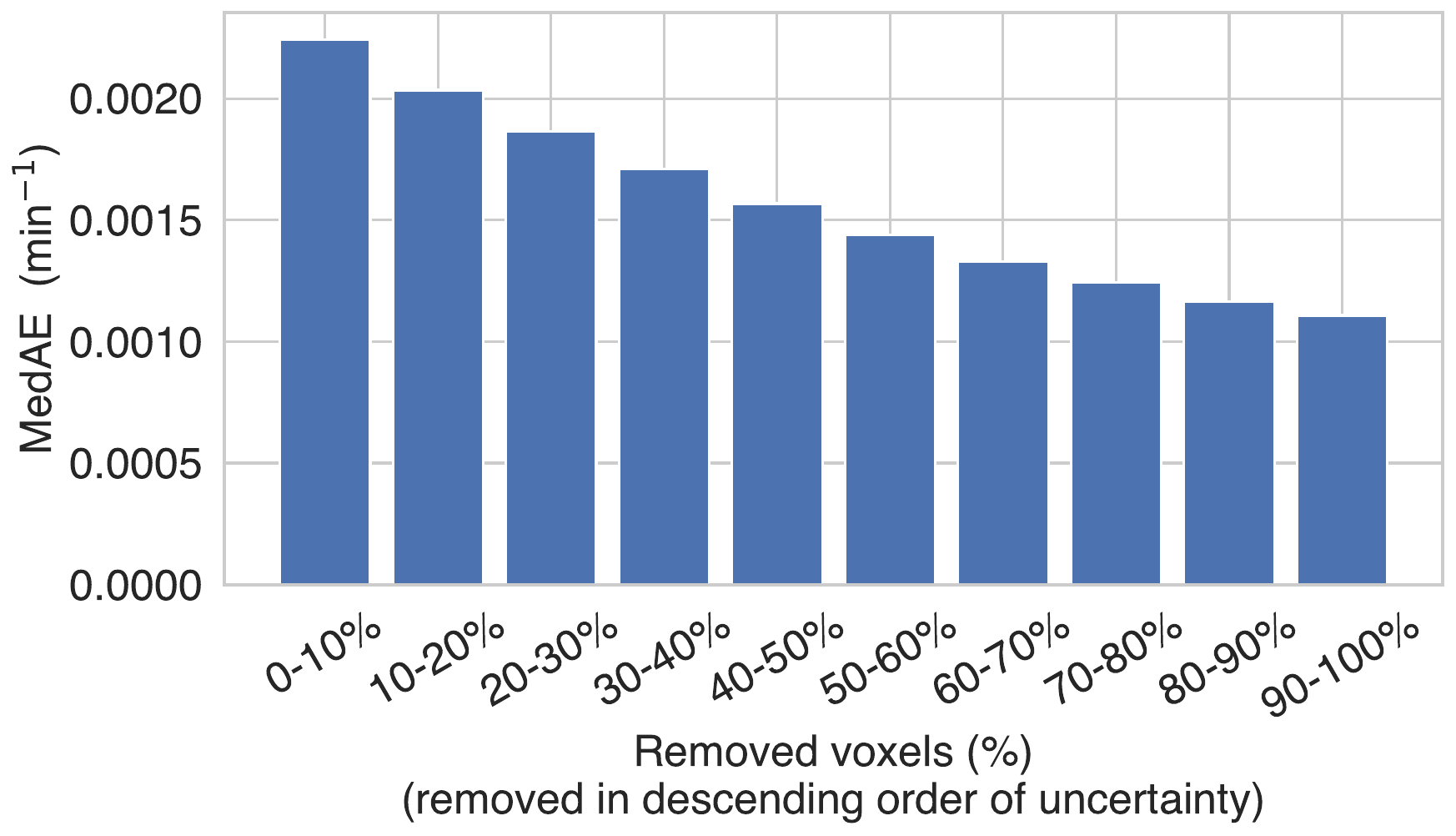}
    \caption{Washout rate ($\lambda_B$) sparsification error plot for all test set voxels. Errors are quantified using the median absolute error (MedAE).}
    \label{fig:sparsification}
\end{figure}

Figure \ref{fig:sparsification} presents the sparsification plot for the uncertainty estimates in the test set. The medAE decreases as voxels are progressively removed in order of decreasing predicted uncertainty. However, it does not reach zero, as it would in an ideal case, suggesting that in some areas, the signal is too weak for the model to properly recognize its uncertainty.

The aleatoric uncertainty was approximately an order of magnitude larger than the epistemic uncertainty, suggesting that the dominant source of uncertainty stems from the inherent noise in the PET data rather than from model instability or poor generalization to the test set.

\subsection{Voxel Type Classification}

\begin{figure}[h]
    \centering
    \includegraphics[width=\linewidth]{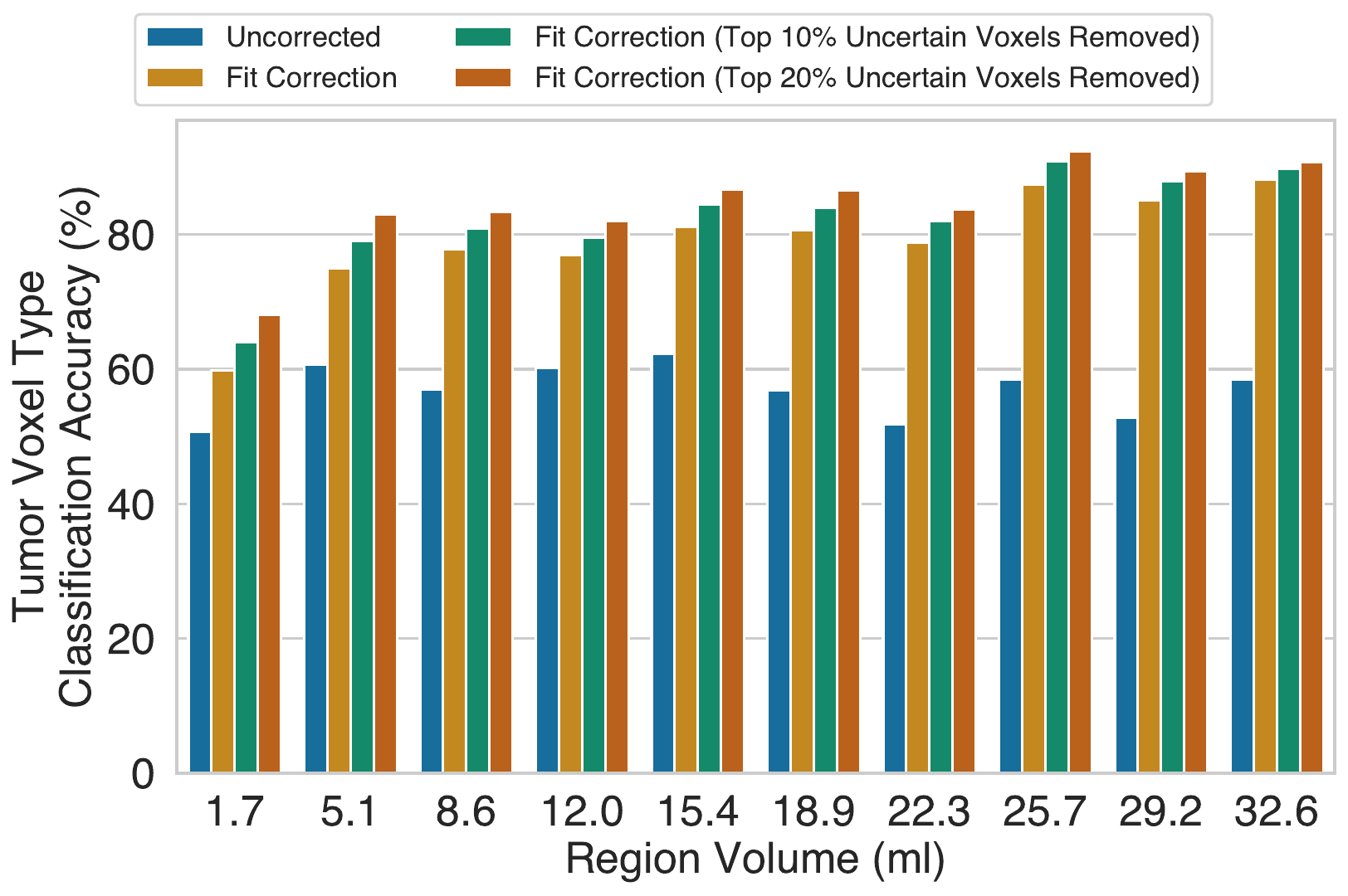}
    \caption{Classification accuracy as a function of region volume when assigning each voxel to one of three vascularity types based on washout rate. Results are presented for maps obtained with the fit correction approach and without correction (Uncorrected). For the corrected maps, the accuracy is additionally reported after excluding the 10\% and 20\% most uncertain voxels.}
    \label{fig:classification-accuracy}
\end{figure}

Figure \ref{fig:classification-accuracy} shows the voxel classification accuracy as a function of region volume, comparing uncorrected and corrected washout rate maps. The corrected maps consistently yield higher classification accuracies across all region volumes,  with the relative improvement increasing for larger regions. We also exclude 10\% and 20\% of voxels with the highest uncertainty in the corrected maps to assess the impact on classification performance. These percentages are selected as exploratory thresholds, representing an implementation strategy that avoids highly uncertain voxels. The classification accuracy improves as voxels with higher uncertainty are removed, as would be expected for well-calibrated uncertainty estimates.

\subsection{Cross-Anatomy Generalization}

Table \ref{tab:liver} presents the performance of the fit correction deep learning model, trained on head-and-neck data, when applied to a liver tumor case. While the improvement is not as pronounced as in the head-and-neck region, likely due to differences in noise characteristics and tumor size, the model still demonstrates substantial improvements in washout rate estimates.

\begin{table}[h]
    \centering
    \caption{Comparison of structural similarity index measure (SSIM) and absolute error (AE) metrics for washout rate maps ($\lambda_B$) obtained with the fit correction approach and without correction (Uncorrected) in a liver tumor. The fit correction model was trained on head-and-neck tumors. Results are reported as median [Q$_1$~--~Q$_3$] for both SSIM (medSSIM~[Q$_1$~--~Q$_3$]) and absolute error (medAE~[Q$_1$~--~Q$_3$]). SSIM values range from -1 to 1 (higher is better); AE values are non-negative (lower is better). The best performance for each metric is indicated in bold.}
    \resizebox{\linewidth}{!}{ 
        \begin{tabular}{lcc}
            \toprule
            Approach (Train/Test Region) & medSSIM [Q$_1$~--~Q$_3$]  & \makecell{medAE [Q$_1$~--~Q$_3$]\\$(10^{-3}\, \text{min}^{-1})$} \\
            \midrule
            \makecell[l]{Uncorrected \\ (Train: -- / Test: Liver)}  & 0.35 (0.31 -- 0.38) & 6.6 (2.7 -- 13.1) \\
            \addlinespace
             \makecell[l]{Fit Correction \\ (Train: Head \& Neck / Test: Liver)} &  \textbf{0.60 (0.58 -- 0.64)} & \textbf{4.3 (1.3 -- 8.0)} \\
            \bottomrule
        \end{tabular}
        }
    \label{tab:liver}
\end{table}

\subsection{Robustness to PET Acquisition Delay After Treatment}

Table \ref{tab:acquisition-timing} presents the performance of the fit correction model, trained on washout rate maps obtained with a delay of 10 minutes, when applied to maps obtained with a delay of 15 minutes. The results remain similar, highlighting the model's robustness to moderate changes in acquisition delay.

\begin{table}[h]
    \centering
    \caption{Comparison of structural similarity index measure (SSIM) and absolute error (AE) metrics for washout rate maps ($\lambda_B$) obtained with the fit correction approach and without correction (Uncorrected) 15 minutes post-treatment, using models trained on maps with either a 15-minute or 10-minute delay. Results are reported as median [Q$_1$~--~Q$_3$] for both SSIM (medSSIM~[Q$_1$~--~Q$_3$]) and absolute error (medAE~[Q$_1$~--~Q$_3$]). SSIM values range from -1 to 1 (higher is better); AE values are non-negative (lower is better). The best performance for each metric is indicated in bold.}
    \resizebox{\linewidth}{!}{ 
        \begin{tabular}{lcc}
            \toprule
            Approach (Train/Test Delay) & medSSIM [Q$_1$~--~Q$_3$]  & \makecell{medAE [Q$_1$-- Q$_3$]\\$(10^{-3}\, \text{min}^{-1})$} \\
            \midrule
            \makecell[l]{Uncorrected \\ (Train: -- / Test: 15 min)} & 0.64 (0.52 -- 0.73) & 5.8 (2.1 -- 14.8) \\
            \makecell[l]{Fit Correction \\ (Train: 15 min / Test: 15 min)} & 0.80 (0.75 -- 0.83)  & \textbf{2.3 (1.0 -- 5.1)} \\
            \makecell[l]{Fit Correction \\ (Train: 10 min / Test: 15 min)}  & \textbf{0.83 (0.78 -- 0.85)}  & 2.4 (1.0 -- 5.3)\\
            \bottomrule
        \end{tabular}
    }
    \label{tab:acquisition-timing}
\end{table}

\subsection{Model Size and Runtime}

The trained model has a size of 254 MB, contains 66 million parameters, and has a computational complexity of 57 GFLOPs. Dataset generation, model training, and inference were all performed on an NVIDIA RTX 4080 SUPER GPU. Generating the dataset for a single scanner required approximately 25 hours. Training the washout rate deep-learning model required 10.8 hours, while training the washed-out deep-learning model required 23 hours, likely due to the increased complexity of the task.

Once trained, the model corrects a washed-out or washout rate map in 0.5 seconds by performing 20 forward passes, each with a different configuration of randomly dropped-out neurons, and averaging the resulting outputs. Each individual inference has a runtime of approximately 25 milliseconds. Fitting the washout parameters to the reconstructed PET frames, which produces the uncorrected maps, is completed in approximately 2.5 seconds.

\section{Discussion}

Visualization of biological washout maps and improved quantitative metrics demonstrate the model's ability to separate intratumoral regions that were previously difficult to distinguish without correction. In addition, the model provides well-calibrated uncertainty estimates, allowing assessment of prediction reliability within a given region~\cite{cbct-uncertainty}.

A previous study investigating $^{15}\text{O}$ ion beam irradiation in rats reported differences in slow biological washout rates between perfused and control tumors, as well as between hypoxic and control tumors, on the order of $9 \times 10^{-3}$ $\text{min}^{-1}$~\cite{washout-vascular}. Following correction with our proposed framework, washout rate errors predominantly fall below this threshold in regions as small as 5.1 mL (see Figure \ref{fig:slow-error-vs-region}). This suggests that, assuming the biological kinetics of $^{11}\text{C}$ are of similar magnitude to those of $^{15}\text{O}$, such tissue-level differences could be detectable at subtumoral scales. 

Even in larger regions, errors in uncorrected washout rate maps remain significant, indicating a strong bias due to spillover effects. This underscores the importance of applying frameworks like the one presented to obtain reliable washout estimates and improve voxel-type classification. To our knowledge, this is the first development of such a framework in the context of post-proton therapy PET. The implementation requires only that the patient be transferred to a PET scanner shortly after treatment and does not involve the injection of additional radiotracers.

Since the models were trained on simulated data from head-and-neck cancer patients with a specific treatment and PET acquisition protocol, evaluating their ability to generalize to maps obtained in different scenarios was important. The fit correction model demonstrated strong generalization to washout rate maps from anatomical regions and PET acquisition timings that differed from those used in training. 

Nonetheless, depending on the specific performance requirements, it may be necessary to redevelop the framework using data more representative of the target patient population and treatment protocols. To enable this, we have open-sourced the complete, flexible framework at \sloppy\href{https://github.com/pcabrales/ppw.git}{https://github.com/pcabrales/ppw.git}.

Improved characterization of biological washout dynamics can increase confidence in assessing tumor status and enable the evaluation of intratumoral heterogeneity. Intratumoral heterogeneity is recognized as an important surrogate biomarker for improved risk stratification and monitoring of tumor progression, reducing the likelihood of under-grading and under-treating tumors and providing a more comprehensive assessment than invasive biopsies~\cite{heterogeneity-signatures, heterogeneity-aggresive, heterogeneity-pet-review}. Intratumoral heterogeneity can help identify patient-specific vulnerabilities, supporting the implementation of personalized treatment strategies~\cite{heterogeneity-personalized}. One such strategy is dose painting, which involves delivering higher doses to aggressive tumor regions, such as hypoxic and necrotic areas~\cite{hypoxia-dose-painting, phase2-dose-painting, hypoxia-pt, sfrt-metabolism, lattice-oxygen}.


Accurate quantification of biological washout is also important given its potential as an indicator of metabolic and vascular response, as well as tumor progression, thereby supporting adaptive treatment strategies~\cite{biomarker-treatment-adaptation, hypoxia-radiation-therapy}. For example, dose-dependent vascular responses have been observed through biological washout, suggesting a possible mechanism for tumor sterilization at higher doses~\cite{radioactive-beam}. 

Moreover, washout mapping may improve understanding of the tumor microenvironment and biological response after FLASH and minibeam radiation therapies, whose underlying mechanisms are not yet fully understood \cite{minibeam, flash}. Since radioactive isotopes are produced directly within the tumor via proton interactions, post-proton therapy PET may provide complementary information compared to conventional PET imaging \cite{fdg-proton-therapy}, where injected radiotracers may not reach certain regions, such as necrotic tumor tissue.

Enhanced biological washout estimates can also improve the simulations used for PET-based dose verification. The current PPW framework builds on previous work that introduced a deep learning and GPU-accelerated workflow for PET-based dose verification, known as PROTOTWIN-PET (PROTOn therapy digital TWIN models for dose verification with PET)~\cite{prototwin-pet}. These two frameworks, PROTOTWIN-PET and PPW, can be simultaneously applied to the same post-treatment PET data, each providing complementary information.




The approaches presented for the PPW framework have been explored in the context of post-proton therapy PET imaging, but they can also be readily adapted for estimating dynamic low-dose PET parameters or analyzing tracer dynamics in small regions~\cite{dyn-pet-denoising}. During training, instead of simulating both proton therapy treatment and PET acquisition, conventional PET scans could be simulated, or real PET data could be undersampled. A variety of training strategies, deep learning considerations, and performance assessments have been introduced, paving the way for further research in this area.

For the medium and fast biological washout components, assessed using the proposed washed-out maps, the improvements were less substantial than for washout rate maps. This is likely due to the additional uncertainty introduced when simulating and obtaining the initial activity without washout. Consequently, and given that washout rates have been more extensively studied and related to tumoral status \cite{washout-vascular, washout-vascular-mri}, the analysis primarily focused on the washout rates. Nevertheless, we consider the washed-out maps a potentially insightful proxy for characterizing the effects of medium and fast washout components. Its investigation is enabled by our simulation pipeline, which is capable of simulating the scenarios without washout. This study serves as proof of concept for the feasibility of estimating washed-out maps and quantifying associated errors. When combined with other metrics, such as washout rate maps or CT-based radiomic features~\cite{ct-texture-heterogeneity}, the washed-out maps may enhance the characterization of intratumoral heterogeneity.

In the washout rate maps, model estimation errors primarily concentrate around the region boundaries, where the estimated boundaries appear less sharp than in the ground truth maps. However, the sharp boundaries in the ground truth maps are artificially sharp because the PET simulation software enforces fixed decay rates within each region. In reality, tumor boundaries are expected to be smoother, as vascular conditions transition gradually rather than abruptly between regions.

Although the sparsification plot in Figure \ref{fig:sparsification} indicates that the uncertainty estimates are well-calibrated, the persistence of errors despite the removal of most high-uncertainty voxels indicates that the uncertainty estimates do not fully capture all regions where the model fails. This limitation is due to the inherently low SNR of PET imaging following proton therapy, a challenge that can't be fully overcome even with a comprehensive framework like the one proposed. While irradiation with heavier ion beams may improve the SNR due to nuclear fragmentation of the projectile ions ~\cite{washout-vascular, offline-bauer}, these therapies are currently less common and more costly~\cite{c-ion-therapy}.

To improve the SNR, it is also possible to minimize the delay between the end of treatment and the start of the PET acquisition. 
Acquiring images with less delay would also make it more challenging to distinguish between contributions from isotopes different from $^{11}$C, such as $^{15}$O and $^{13}$N. Nevertheless, regional and treatment-induced differences in tumor washout dynamics could likely still be observed and prove informative when analyzing an effective decay signal that combines the contributions of multiple isotopes~\cite{kira-15o}, even if their individual washout components cannot be fully disentangled.



Validation through phantom studies and clinical data is essential and is currently being pursued. Our goal is to determine whether washout estimates are reliable and to use dynamic contrast-enhanced MRI (DCE-MRI) alongside other imaging techniques to assess whether different regions consistently show distinct characteristics across modalities. Moreover, we aim to investigate whether washout patterns and changes throughout treatment correlate with outcomes. While our current analysis focuses on the tumor, washout changes in surrounding tissues may also reveal post-treatment impairments~\cite{white-matter-reduced-diffusion}.

\section{Conclusions}

The framework presented in this study effectively leverages the denoising and resolution recovery capabilities of deep learning to estimate voxel-wise biological washout parameters with associated uncertainty from post-proton therapy PET imaging. Trained on Monte Carlo simulations of cancer patients' digital twins, the models help capture biological washout dynamics and heterogeneities within tumors. These non-invasive assessments have potential applications in monitoring treatment progression, guiding treatment adaptation, and enhancing dose verification models, ultimately supporting more personalized therapies. The proposed approach requires neither specialized equipment nor modifications to existing hardware or software, enabling its implementation at any proton therapy facility equipped with a PET scanner.

\section*{Acknowledgments}
This work was supported by Spanish MCIN/AEI/10.13039/501100011033 and the European Union NextGenerationEU/PRTR under grants TED2021-130592B-I00 PROTOTWIN, PID2021-126998OB-I00 FASCINA, and PDC2022-133057-I00 3PET; by Comunidad de Madrid (Spain) under projects ASAP-CM P2022/BMD-7434 and TAU-CM PR47/21 (PRTR), and by Grupo de Fisica Nuclear (910059) at Universidad Complutense de Madrid. Additionally, Pablo Cabrales acknowledges support from the Complutense University Predoctoral Fellowship (CT15-23), funded by the Universidad Complutense de Madrid and Banco Santander, as well as from the \textit{Ayuda para estancias breves} fellowship (EB33/24), funded by the Universidad Complutense de Madrid. M.Pérez-Liva was supported by the Program Ramón y Cajal RYC2021-032739-I, MCIN/AEI/10.13039/501100011033, by the European Union NextGenerationEU/PRTR, by INVENTOR PID2022-137114OA-I00, and by LUNABRAIN (PR37/24 TEC-2024/TEC-43).

\section*{Data and Code Availability}
The CT images and radiation therapy structures used in this study are from the Head-Neck-PET-CT dataset hosted by The Cancer Imaging Archive (TCIA)~\cite{head-dataset}. Because the images include facial anatomy that could permit re-identification, the dataset is distributed under the TCIA Restricted License. Researchers can obtain the dataset by creating a free TCIA user account and requesting access. Approval is granted by TCIA, not by the authors, and the authors had no special access privileges. 

The use of these anonymized data for research purposes was approved by the Institutional Review Boards (IRBs) of all participating institutions. Retrospective analyses were conducted in compliance with applicable guidelines and regulations, outlined by the ethical committees of the respective institutions under protocol number MM-JGH-CR15-50. 

To support further research, all code used for data preprocessing, treatment planning, simulation, and deep learning model training and validation, along with corresponding usage instructions, is available at \sloppy\href{https://github.com/pcabrales/ppw.git}{https://github.com/pcabrales/ppw.git}.




\bibliographystyle{elsarticle-num} 
\bibliography{references}

\end{document}